\documentstyle[12pt]{article}
\def\la{\mathrel{\mathpalette\fun <}}

\def\fun#1#2{\lower3.6pt\vbox{\baselineskip0pt\lineskip.9pt
\ialign{$\mathsurround=0pt#1\hfil##\hfil$\crcr#2\crcr\sim\crcr}}}

\begin{document}
\topmargin -0.5cm
\oddsidemargin -0.3cm

\begin{titlepage}
\pagestyle{empty}
\begin{flushright}
{ITEP 43-96}
\end{flushright}
\vspace*{1mm}

\begin{center}
{\large\bf LEPTONIC PHOTON AND
 LIGHT}
\vspace{5mm}
{\large\bf ELEMENT ABUNDANCIES}
\vspace{2cm}

{\bf L.B.  Okun}\\
\vspace{0.1in}
ITEP, 117218, Moscow, Russia

\vspace{0.5in}
 {\bf Abstract \\}
 \end{center}

In the framework of a model, in which a single leptonic photon $\gamma_l$
has the same coupling to the doublets $\mu\nu_{\mu}$ and
$\bar{\tau}\bar{\nu}_{\tau}$, there is no cosmological bound on the
strength of this coupling.

\end{titlepage}

\vfill\eject
\pagestyle{empty}

\setcounter{page}{1}
\pagestyle{plain}

A few papers devoted to hypothetical leptonic photons have appeared recently
\cite{1,2,3,4} (for the first discussion of leptonic photons see ref.
\cite{5} for that of baryonic photons -- ref. \cite{6}).

In particular in ref. \cite{1} an anomaly-free model was suggested in which
a single leptonic photon $\gamma_l$ had the same coupling to the doublets
$\mu\nu_{\mu}$ and $\bar{\nu}_{\tau}\bar{\tau}$ and was decoupled from
$e\nu_e$. In this note we will consider this model from the point o
f view
of cosmology.

The cosmological limit set by the theory of nucleosynthesis and the
present data on abundancies of light elements  allows  the
existence in equilibrium at $T\sim 1$ MeV of only one new
light particle with two degrees of freedom \cite{7, 8, 9} in addition
to three left-handed neutrinos ($\nu_e$, $\nu_{\mu}$, $\nu_{\tau}$),
their right-handed antineutrinos and the ordinary photon (for the
first estimates see \cite{10, 11}).

In the framework of our model the Dirac masses of mu
onic and tauonic
neutrinos are forbidden if the leptonic photon coupling is strong
enough ($\alpha_{\mu} \equiv \alpha_{\tau}\equiv \alpha_e >
10^{-11}$). This follows from considerations of ref. \cite{4}. Due to
strict conservation of muonic and tauonic charges the right-handed
$\nu_{\mu}$ and $\nu_{\tau}$ and the left-handed $\bar{\nu}_{\mu}$
and $\bar{\nu}_{\tau}$ must have the same strength of the coupling
with $\gamma_l$, as the ordinary left-handed $\nu_{\mu}$ and
$\nu_{\tau}$ and right-handed
 $\bar{\nu}_{\mu}$ and
$\bar{\nu}_{\tau}$. As a result all these neutrinos would be in
thermal equilibrium at the moment of nucleosynthesis and there would
exist 3 extra particles with two degrees of freedom each, while the
cosmological limit allows only one such particle.

Thus the $\nu_{\mu}$ and $\nu_{\tau}$ must be either
two-component and hence
massless or have a
common non-diagonal Majorana mass of ZKM type \cite{12, 13}.
In the former case the theory
would be plagued by infrared divergence
s, which however are not so
damaging, as to make it non-viable. In the latter case a $\tau^+$
lepton may be produced by an originally clean $\nu_{\mu}$-beam, but
the probability of such event would be negligibly small for small
values of the mass.

If further progress in understanding the nucleosynthesis excludes the
existence of any extra light particle in equilibrium
at $T\sim 1$ MeV, then the leptonic
photon $\gamma_l$
is allowed to come into equilibrium with other light particles only
at $T <
 1$ MeV. This would set an upper limit on the leptonic fine
structure constant $\alpha_l \equiv \alpha_{\tau} \equiv \alpha_{\mu}$. By
comparing the rate of reactions
$\nu_{\mu}\bar{\nu}_{\mu} \leftrightarrow \gamma_l
\gamma_l$, $\nu_{\tau}\bar{\nu}_{\tau} \leftrightarrow \gamma_l
\gamma_l$, proportional to $\alpha_l^2 T$, with the Hubble expansion
rate, proportional to $T^2/M_{pl}$, at $T < 1$ MeV, as was done in
ref. \cite{4}, one would get, $\alpha_l \la 10^{-11}$.

But with the present data on 
the abundancies of light elements there
is no cosmological limit on $\alpha_l$ in the framework of the model
under discussion.

As already mentioned elsewhere \cite{1, 5}, the present experimental
upper limit on $\alpha_{\mu}$ is $10^{-5}\alpha$. It may be improved
by an order of magnitude \cite{14} by analyzing the data of high
energy neutrino experiments.

Note that if the model under discussion is correct, there should be
no $\nu_{\mu}$ -- $\nu_{\tau}$ oscillations in experiments such as
CHORU
S and NOMAD. Note also that with $\alpha_{\tau}$ of the order of
$10^{-5}$ there would be no manifestations of virtual tauonic photon
in the decays of $Z$-bosons:
$Z\to \nu_{\tau}\bar{\nu}_{\tau}(\tau\bar{\tau})$,
$Z\to \tau\bar{\tau}(\nu_{\tau}\bar{\nu}_{\tau})$,
where brackets indicate a narrow Dalitz-like pair. Still search for
such decays is of great interest.

This research was made possible in part by grants 93-02-14431 of the Russian
Foundation for Basic Research. I am grateful to A.D.Dolgov
, V.A.Novikov,
A.N.Rozanov, M.B.Voloshin and M.I.Vysotsky for helpful discussion.

\end{document}